\documentclass[twocolumn,openany,openright,amsmath,amssymb,superscriptaddress]{revtex4-1}

\usepackage{float}
\usepackage{latexsym} 
\usepackage{amsmath,amsthm}
\usepackage{ifpdf}
\usepackage{epstopdf}
\usepackage{subfig}
\usepackage{soul}
\usepackage{dcolumn}
\usepackage{bm}
\usepackage{braket}
\usepackage{wrapfig}
\usepackage[dvipsnames]{xcolor}
\usepackage{color}
\usepackage{hyperref}

\usepackage{graphicx}

\begin{document}
\newcommand{\umlaut}[1]{\ddot{\textrm{#1}}}
\newcommand{\beq}{\begin{equation}}
\newcommand{\eeq}{\end{equation}}
\newcommand{\barr}{\begin{eqnarray}}
\newcommand{\earr}{\end{eqnarray}}
\newcommand{\bseq}{\begin{subequations}}
\newcommand{\eseq}{\end{subequations}}
\newcommand{\oper}[1]{\hat{#1}}
\newcommand{\adj}[1]{\oper{#1}^{\dagger}}
\newcommand{\ann}{\hat{a}}
\newcommand{\creat}{\hat{a}^{\dagger}}
\newcommand{\commutator}[2]{[\oper{#1},\oper{#2}]}
\newcommand{\outerP}[2]{|#1\rangle\langle#2|}
\newcommand{\expectation}[3]{\langle #1|#2|#3\rangle}
\newcommand{\proiettore}[1]{\ket{#1}\bra{#1}}
\newcommand{\closure}[1]{\sum_{#1}\proiettore{#1}=1}
\newcommand{\closureInt}[1]{\int d#1\proiettore{#1}=1}
\newcommand{\vett}[1]{\textbf{#1}}
\newcommand{\uvett}[1]{\hat{\textbf{#1}}}
\newcommand{\derivata}[2]{\frac{\partial #1}{\partial #2}}

\newcommand{\creak}[1]{\hat{a}^{\dagger}_{#1}}
\newcommand{\annak}[1]{\hat{a}_{#1}}
\newcommand{\crebk}[1]{\hat{b}^{\dagger}_{#1}}
\newcommand{\annbk}[1]{\hat{b}_{#1}}

\newcommand{\crek}[2]{\hat{#1}^{\dagger}_{#2}}
\newcommand{\annk}[2]{\hat{#1}_{#2}}
\newcommand{\red}[1]{{\color{red}{#1}}}
\newcommand{\blue}[1]{{\color{blue}{#1}}}

\title{Structured Light-Matter Interaction: Twisted Photons in Graphene}
\author{Yaraslau Tamashevich}
\affiliation{Faculty of Engineering and Natural Sciences, Tampere University, Tampere, Finland}
\author{Marco Ornigotti}
\affiliation{Faculty of Engineering and Natural Sciences, Tampere University, Tampere, Finland}

\begin{abstract}
In this work we present a numerical framework for studying the interaction of structured electromagnetic fields, i.e., light pulses carrying orbital angular momentum (OAM), interacting with a single layer of graphene. Our approach is based on a two-step process, where first the interaction of structured light with matter, modelled by a suitable Dirac equation is solved and then coupled with Maxwell's equations, for studying the properties in both space and frequency domain of the generated nonlinear electromagnetic field. As an example of application of our method we investigate third harmonic generation by an OAM pulse. We check that OAM conservation through harmonic generation is conserved, and we report on both the spatial and frequency features of the third harmonic signal, comparing it with the traditional frequency response for spatially uniform field.
\end{abstract}

\maketitle

\section{Introduction}
Since the discovery of graphene \cite{novoselov_electric_2004} in 2004, a lot of effort in theoretical and experimental works has been done to unravel, investigate and control its peculiar properties, such as minimal conductivity \cite{refCond1}, universal absorbance \cite{univAbs1}, anomalous Quantum Hall Effect \cite{novoselov_qhe,zhang_qhe}, and high nonlinear optical response \cite{wright_strong_2009,hafez_terahertz_2020}, to name a few. 
Inspired by the new physics graphene brought forward, the last decades saw the dawn of a new class of materials, consisting on only one or few atomic layers, the so-called 2D materials, of which transition metal dichalcogenides (TMDs)   \cite{manzeli_2d_2017,autere_nonlinear_2018},  hexagonal boron nitride (hBN) \cite{caldwellphotonics2019,neto2009electronic},  and black phosphorous \cite{blackPh} are amongst the most interesting for Photonics applications.

The most interesting characteristic these 2D materials possess, that make them very interesting and promising for applications in different areas of classical and quantum Photonics is their high nonlinear response, which has been reported being several orders of magnitude higher, than their bulk counterpart. Several experiments have, in fact, reported second- \cite{zeng2013optical} and third-harmonic generation \cite{kumar2013third}, four-wave mixing \cite{hendry_coherent_2010}, and optical limiting \cite{dong_optical_2015} from graphene or other 2D materials, allowing for a new generation of integrated photonic devices, that can work in a low-power, energy-efficient regime, which enables a new generation of photonic technologies.

While graphene is intrinsically centrosymmetric, and therefore lacks second order response, recent works have demonstrated how it is possible to enable second-harmonic generation in graphene by stress- or strain-induced magnetic fields \cite{graphMagField1, ornigotti_strain_2021}, or breaking centrosymmetry via extreme light-matter interaction \cite{fabio1,graphMagField2}. 

On a seemingly parallel trail, structured light, i.e., electromagnetic wave with a nonuniform intensity and polarisation distribution \cite{OAMbook}, has been widely studied, since the seminal works by Berry and Nye in 1974 \cite{berryNye} and Allen and Woerdman in 1992 \cite{allenWoerdman}, and today it represents a versatile tool used in several different research laboratories and industrial environments alike, showing a wide range of applications in different areas of physics, ranging from particle manipulation \cite{oam1}, microscopy \cite{oam3}, spectroscopy \cite{oam4}, biophysics \cite{bioOAM}, classical \cite{oam5} and quantum \cite{oam6} communications, quantum information \cite{oam2}, and material processing \cite{materialOAM}, to name a few.
%

Despite the considerable amount of research carried out in both these very active and impactful research fields, investigating the benefits and potential new physics that would result by combining these two disciplines is still a mostly unexplored research direction. From a theoretical perspective, progress in coupling structured light with matter has resulted in the twisted gauge formalism \cite{twistedGauge}, and the twisted electrodynamics framework \cite{qedTwisted}, with the former having been applied to semiconductors \cite{semicond1,semicond2, semicond3} and to ring-shaped monolayers of graphene \cite{farias_photoexcitation_2013} and graphene-enhanced nanostructures \cite{grapheneMeta}, and, recently,light carrying orbital angular momentum (OAM) has been also proposed as a probe to investigate centrifugal photovoltaic and photogalvanic effect in semiconductors \cite{watzel_centrifugal_2016}. On the experimental side, vector beams have been used to drive and control frequency generation in gapped semiconductors \cite{ebrahim1}, and as a probe for topological phase transition of matter through high-harmonic spectroscopy \cite{garcia2024topological}. Moreover, graphene has also been proposed as a platform to generate polarization-entangled photons exploiting the topological phase naturally encoded in the Dirac cones appearing in its low-energy band structure \cite{entGraphene}. All these examples certainly highlight the interest of the community to study the interaction of structured light with exotic materials, such as graphene or other 2D or 2D-like materials. However, the framework used for all these works is very different, and frequently ad-hoc adapted to the specific needs of the particular problem.

%

In this work, we introduce a framework to describe the interaction of structured light with graphene, which combines both analytical and numerical techniques to provide a versatile and reliable tool for this growing research field. Our approach is based on a two-step process, where first the interaction of structured light with matter, modelled by a suitable Dirac equation (as it is frequently the case for 2D materials), is solved and then coupled, in a second step, with Maxwell's equations, for studying the properties, in both space and frequency domain, of the light generated by the linear and nonlinear interaction of structured light with matter. Although the results and methods presented here focus on graphene, our framework can be easily extended any 2D material, by just replacing the mass and dispersion relation of graphene to that of the specific material considered.

This work is organised as follows: in Sect. II, we derive the low-energy Hamiltonian of graphene, and briefly discuss possible generalisations to other materials. Section III introduces the interaction with light through minimal coupling, sets up the Dirac-like equation of motion for the electron wave function, and sets up the electromagnetic problem, using the current generated from the interaction of graphene with the impinging electromagnetic pulse as a source for Maxwell's equations describing the interaction-induced electromagnetic field, i.e., nonlinear response of the material. The results of our simulations are then displayed and discussed in Sect. IV, and conclusions are drawn in Sect. V.

\begin{figure}[!t]
\begin{center}\label{graphene_band}
\includegraphics[width=0.4\textwidth]{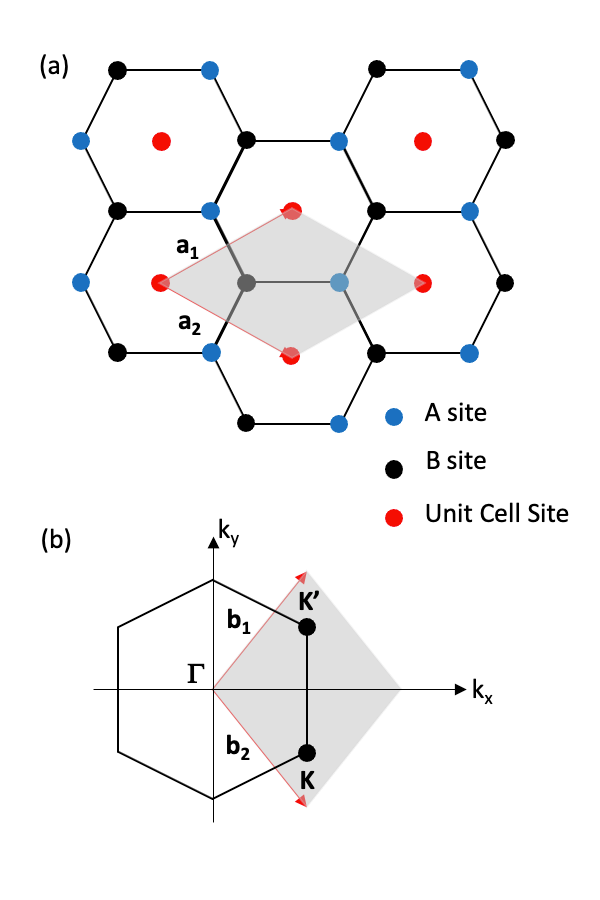}
\caption{(a) Lattice structure of graphene. The unit cell, containing two atoms (one belonging to lattice A, the other to lattice B), is shaded in gray. The basis vectors used in this work are chosen to be $\vett{a}_1=(a/2)(3,\sqrt{3})$, and $\vett{a}_2=(a/2)(3,-\sqrt{3})$. (b) Reciprocal space lattice structure of graphene (the Wigner-Seitz cell is shaded in grey). The reciprocal lattice basis vectors are defined as $\vett{b}_1=(2\pi/3a)(1,\sqrt{3})$ and $\vett{b}_2=(2\pi/3a)(1,-\sqrt{3})$.The two inequivalent Dirac points $\vett{K}$ and $\vett{K'}$ are defined as $\vett{K}=(2\pi/3a)(1,-1/\sqrt{3})$, and $\vett{K'}=(2\pi/3a)(1,1/\sqrt{3})$,respectively.}
\label{figure1}
\end{center}
\end{figure}
\section{Low-Energy Hamiltonian for Graphene}\label{sec:graphene}
%
The aim of this section is to derive the Dirac-like equation for massless electrons in the vicinity of a Dirac point in graphene. To do so, we need first to derive the low energy Hamiltonian for graphene. The crystalline Hamiltonian for electrons in graphene reads
\barr\label{eq4}
\hat{H}_0&=&\int\,d^3r\,\hat{\psi}^{\dagger}(\vett{r})\left[\frac{\hat{\vett{k}}^2}{2m}+\phi_c(\vett{r})+V_B(\vett{r})\right]\hat{\psi}(\vett{r}) \\  \nonumber &\equiv&\int\,d^3r\,\hat{\psi}^{\dagger}(\vett{r})\mathcal{H}_0\hat{\psi}(\vett{r}),
\earr
where $\phi_c(\vett{r})$ is the Coulomb potential, $V_B(\vett{r})$ is the crystalline (periodic) potential, and $\hat{\psi}(\vett{r})$ is the electron wave function operator. The interaction with the external electromagnetic field can either described with the usual electric-dipole interaction Hamiltonian
%
\beq\label{eq4bis}
\hat{H}_I=-e\int\,d^3r\,\hat{\psi}^{\dagger}(\vett{r})\,\left[\vett{r}\cdot\vett{E}_T(\vett{r},t)\right]\,\hat{\psi}(\vett{r}),
\eeq
or by means of minimal coupling, by replacing the kinetic momentum $\vett{k}$ with the canonical momentum $(\vett{k}+e\vett{A})$. IN this work, and, in particular, in the next section, we use the latter approach, together with the Coulomb gauge ($\nabla\cdot\vett{A}=0$), to describe the interaction of light with graphene.

To calculate the graphene Hamiltonian, we use the following Bloch representation for the electron wavefunction operator
\barr
\label{eq6}
\hat{\psi}(\vett{r})=\frac{1}{\sqrt{N}}\sum_{k}\sum_{n=0}^Ne^{i\vett{k}\cdot\vett{R}_n}\Big[\phi_A(\vett{r}-\vett{R}_n)\annak{k} \\ \nonumber
+\phi_B(\vett{r}-\vett{R}_n)\annbk{k}\Big],
\earr
where $\phi_{A,B}(\vett{r})$ are $2p_z$-orbital wavefunctions at site $A$ and $B$, respectively,  $\vett{R}_n=j\vett{a}_1+m\vett{a}_2$ is a Bravais lattice vector (see Fig. \ref{figure1}), and $\annak{k},\annbk{k}$ are the electron site operators, annihilating an electron on site A or B, respectively.
Let us focus on the free Hamiltonian. By substituting Eq. \eqref{eq6} into  Eq. \eqref{eq4}, it is not difficult to see, that the free Hamiltonian $\hat{H}_0$ can be then reduced to 
\beq
\hat{H}_0=\gamma\sum_k\left[f(\vett{k})\creak{k}\annbk{k}+\text{h.c.}\right],
\eeq
where $\gamma$ is the hopping amplitude  (defined as overlap integral between the orbital wave functions of the two sites) and $f(\vett{k})=1+\exp{(i\vett{k}\cdot\vett{a}_1)}+\exp{(i\vett{k}\cdot\vett{a}_2)}$, and $\text{h.c.}$ stands for Hermitian conjugate. Diagonalising the above free Hamiltonian leads to the band structure (eigenvalues) of graphene, i.e., $E(\vett{k})=\pm\gamma|f(\vett{k})|$, and its corresponding eigenvectors
\beq\label{eq18}
u_{\pm}(\vett{k})=\frac{1}{\sqrt{2}}\left(\begin{array}{c}
e^{i\frac{\varphi(\vett{k})}{2}}\\
\pm\,e^{-i\frac{\varphi(\vett{k}}{2})}
\end{array}\right),
\eeq
where $\varphi(\vett{k})=\text{arg}\left[f(\vett{k})\right]$. A plot of the full band structure of graphene is shown in Fig. \ref{graphene_band}.

We can transform the site operators as well into the diagonal basis of the Hamiltonian $\hat{H}_0$, obtaining
\beq\label{eq20}
\left(\begin{array}{c}
\annk{e}{k}\\
\crek{h}{-k}
\end{array}\right)=\frac{1}{\sqrt{2}}\left(\begin{array}{cc}
e^{-i\frac{\varphi(\vett{k})}{2}} & e^{i\frac{\varphi(\vett{k})}{2}}\\
e^{-i\frac{\varphi(\vett{k})}{2}} & -e^{i\frac{\varphi(\vett{k})}{2}}
\end{array}\right)\left(\begin{array}{c}
\annk{a}{k}\\
\annk{b}{k}
\end{array}\right).
\eeq
Notice that this is equivalent to the electron-hole representation discussed, for example, in Ref. \cite{koch}, where $\annk{e}{k}$ annihilates an electron with momentum $k$, and $\crek{h}{-k}$ creates a hole with momentum $-k$.

\begin{figure}
\begin{center}
\includegraphics[width=0.45\textwidth]{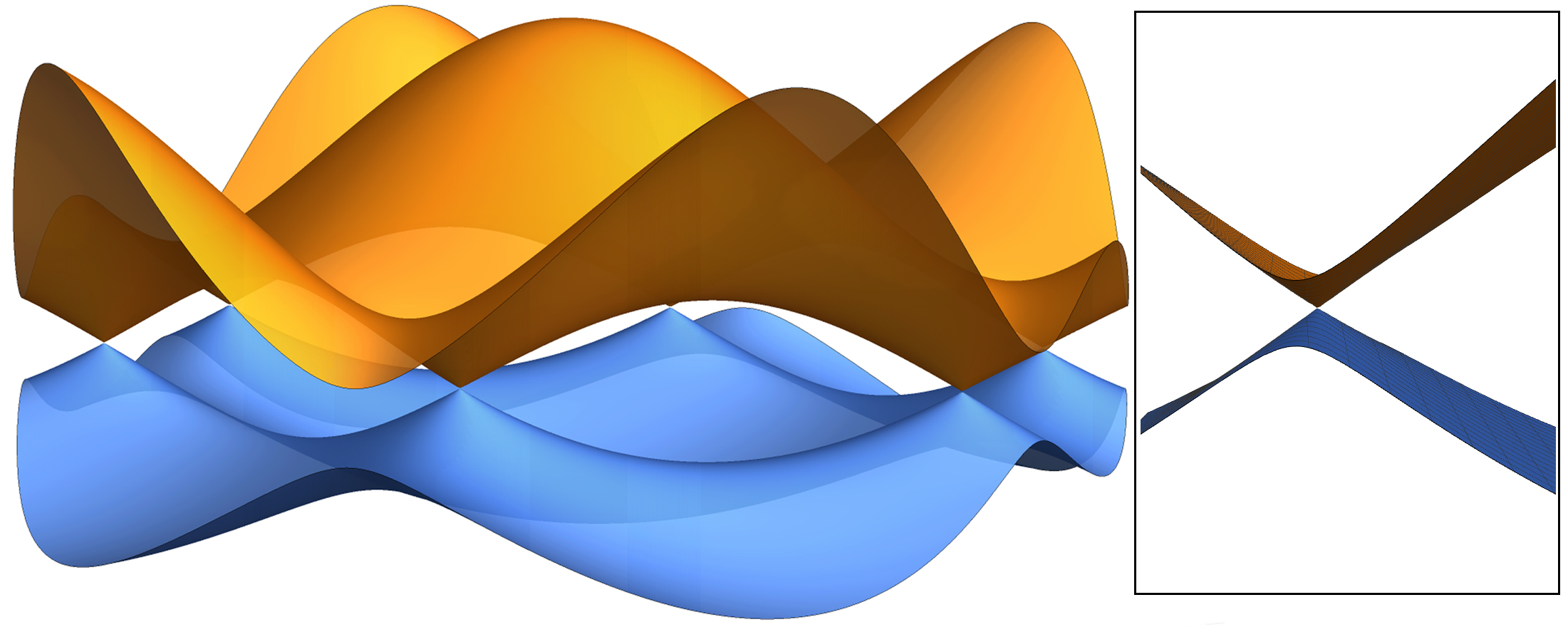}
\caption{Band structure of graphene. You can see 6 touching point of valence (orange) and conduction (blue) band which are called Dirac points. Locally Dirac point can be represented as a touching of two identical cones (see figure inside rectangular) that allows there treating as a massless particles.}
\label{graphene_band}
\end{center}
\end{figure}
The free Hamiltonian in the electron-hole representation can be then written as follows
\beq
\hat{H}_0=\sum_k\,E(\vett{k})\left(\crek{e}{k}\annk{e}{k}+\crek{h}{-k}\annk{h}{-k}\right).
\eeq

\subsection{Low-Energy Approximation}
We can expand the eigenvalues $E(\vett{k})$ around the zeros of $f(\vett{k})$, which happen to be the two inequivalent Dirac points $\vett{K}$ and $\vett{K'}$ [see Fig. \ref{figure1} (b)] \cite{katsnelson} to obtain the low-energy approximation of $\hat{H}_0$ around the Dirac points. To do that, we first redefine the coordinates of the Dirac points as $\vett{K}=(2\pi/3a)(1,\xi/\sqrt{3})$, where $\xi=\pm 1$ is the so-called valley index, and then expand $f(\vett{k})$ around $\vett{q}=\vett{k}+\vett{K}$ obtaining
\barr
f(\vett{q})&\simeq& f(\vett{K})+\nabla_kf(\vett{k})\Big|_{\vett{K}}\vett{q}\nonumber\\
&=&\frac{3ai}{2}\left(k_x-i\xi\,k_y\right)\nonumber\\
\earr
Substituting this result in the expression of the Hamiltonian $\hat{H}_0$ and defining the Fermi velocity as $v_f = 3a\gamma/2\hbar\simeq c/300$, we get the familiar form of the low-energy graphene Hamiltonian in $k$-space, i.e.,
\beq
\hat{H}(\vett{k})=\hbar v_f(\sigma_xk_x+\xi\sigma_y k_y),
\eeq
where $\xi=\pm1$ is the valley index, and $\sigma_i$ are Pauli matrices. The eigenvalues of $\hat{H}(\vett{k})$ are now simply $E(\vett{k})=\hbar v_f|\vett{k}|$. Taking the Fourier transform in space of the above Hamiltonian leads to its representation in direct space, i.e.,
\beq\label{free_hamiltonian}
\hat{H}=\hbar v_f(\sigma_x \partial_{x} +\xi \sigma_y \partial_{y}),
\eeq
which is more suitable for the structured-light-matter interaction model we want to build in this work, since it allows to directly input the spatial intensity profile of structured light pulses, avoiding the computationally heavy task of describing the interaction of matter with $k-$dependent fields in Fourier space \cite{quinteiro1}.


%
\section{Interaction Hamiltonian}
As discussed in the previous section, to include interaction of graphene with light, we could use minimal coupling and replace the kinetic momentum $\vett{k}$ in Eq. \eqref{eq4} with the canonical momentum $(\vett{k}+e\vett{A})$, and redo all the calculations above to include the interaction term. Equivalently, we can simply make this substitution in Eq. \eqref{free_hamiltonian} to obtain
%
\beq\label{interaction_hamiltonian}
\hat{H}=\hbar v_f \left[\sigma_x (k_{x} + \frac{e}{\hbar} A_{x}) +\xi \sigma_y (k_{y} + \frac{e}{\hbar} A_{y}) \right].
\eeq
The equations of motion generated by this Hamiltonian are in the form of a massless Dirac equation, given by
\begin{equation}\label{dirac_equation}
    i \hbar \partial_{t} \psi =  \hat{H} \psi. 
\end{equation}
This equation describes the dynamics of spin $1/2$ particles, represented in general by the two-component spinor $\psi = (\psi_{1}, \psi_{2})$, with $\psi_{1,2}$ being the electron (hole) wave function in valence and conduction bands, interacting with an electromagnetic field described by the vector potential $\vett{A}(\vett{r},t)=A_x(\vett{r},t)\uvett{x}+A_y(\vett{r},t)\uvett{y}\equiv A_x\uvett{x}+A_y\uvett{y}$. Equation \eqref{dirac_equation} can be rewritten as a system of two coupled evolution equations for the electron and hole wave functions $\psi_{1,2}$ as
\begin{subequations}\label{dirac_system}
\begin{align}
i \partial_{t} \psi_{1} &= v_{f}  \left( \partial_{x}-  i \xi \partial_{y} \right) \psi_{2}  + \frac{e v_f}{\hbar}\left( A_{x} - i \xi A_{y} \right) \psi_{2},\\
i \partial_{t} \psi_{2} &= v_{f} \left( \partial_{x} +  i \xi \partial_{y} \right)  \psi_{1}  + \frac{e v_f}{\hbar} \left( A_{x} + i \xi A_{y} \right) \psi_{1}.
\end{align}
\end{subequations}
The above equation describes exactly the light-matter interaction within the low energy approximation of graphene, i.e., for light pulses with angular spectra well-localised around one or both Dirac valleys. Unfortunately, a general analytical solution for the above equation, for an arbitrary spatiotemporal shape for the impinging electromagnetic field, is not available, although some interesting special cases have been discussed \cite{Wolkow1935berEK}. 

Even though an exact, analytical solution of Eqs. \eqref{dirac_system} is not available, setting up a suitable solver for it will give us the possibility to explore the linear and, most interestingly, nonlinear interaction of graphene with nonconventional electromagnetic fields, with nonstandard spatial and temporal shapes. This is, essentially, the aim of this work, and Eqs. \eqref{dirac_system} constitute the core of the framework we will be using in the next section to solve them numerically.

An important quantity that can be derived directly from Eqs. \eqref{dirac_system}, which is of particular usefulness for calculating the nonlinear response of graphene, is the space- and time-dependent current, obtained from the solutions $\psi_{1,2}$ of Eqs. \eqref{dirac_system} as follows \cite{srednicki}
\begin{equation}\label{current}
\vett{J}_{Dirac}(\vett{r},t)=2\sum_{\xi}\left[\operatorname{Re}\{\psi_1\psi_2^*\}\,\uvett{x}+\operatorname{Im}\{\psi_1\psi_2^*\}\,\uvett{y}\right].
\end{equation}
The current defined above contains all the information about the linear and nonlinear response of the graphene monolayer, to the impinging electromagnetic field. In particular, its Fourier transform in time grants access to the frequency response and, ultimately, to the harmonic spectrum \cite{miaReview2023}. The (position-dependent) spectrum of emitted radiation $S(\vett{r},\omega)$, in fact, is given by \cite{ishikawa}
\beq
S(\vett{r},\omega)=|\omega\,\tilde{\vett{J}}_{Dirac}(\vett{r},\omega)|^2,
\eeq
where $\tilde{\vett{J}}_{Dirac}(\vett{r},\omega)$ is the Fourier transform of the current defined in Eq. \eqref{current}. Integrating the above quantity in space gives then access to the spectrum of emitted radiation $S(\omega)$, from which the nonlinear response of graphene can be deduced.

In general, however, only the time dependence of the vector potential in Eq. \eqref{dirac_system} is taken into account, since it is frequently assumed that the electromagnetic field has uniform intensity. Here, as it is shown below, we instead consider the case where the electromagnetic field has a nontrivial spatial distribution of intensity, so that the generated current will be time- and space-dependent, thus giving access to a new regime of operation and new interplay between spatial and spectral degrees of freedom of the field. To gain access to these features we adopt a simple strategy: we encode the particular spatial structure of the impinging electromagnetic field in its vector potential, and use it to solve  Eqs. \eqref{dirac_system}, getting then access to $\psi_{1,2}$. We then use this result to calculate the induced Dirac current with Eq. \eqref{current}, and look at either the optical response $S(\omega)$ or the spatial distribution of the current. For accessing the spatial information of the (nonlinear) electormagnetic field generated by the Dirac current both in the near- and far-field, we couple Eqs. \eqref{dirac_system} with Maxwell's equations, using the Dirac current as a source for the wave equation, i.e.,
\beq\label{eqMaxwell}
\left(\nabla^2-\frac{1}{c^2}\frac{\partial^2}{\partial t^2}\right)\vett{E}=\mu_0\frac{\partial\vett{J}_{Dirac}}{\partial t}.
\eeq
In this way, we have full access to the spatial properties of the generated nonlinear electromagnetic field.
\begin{figure}
\begin{center}
\includegraphics[width=0.45\textwidth]{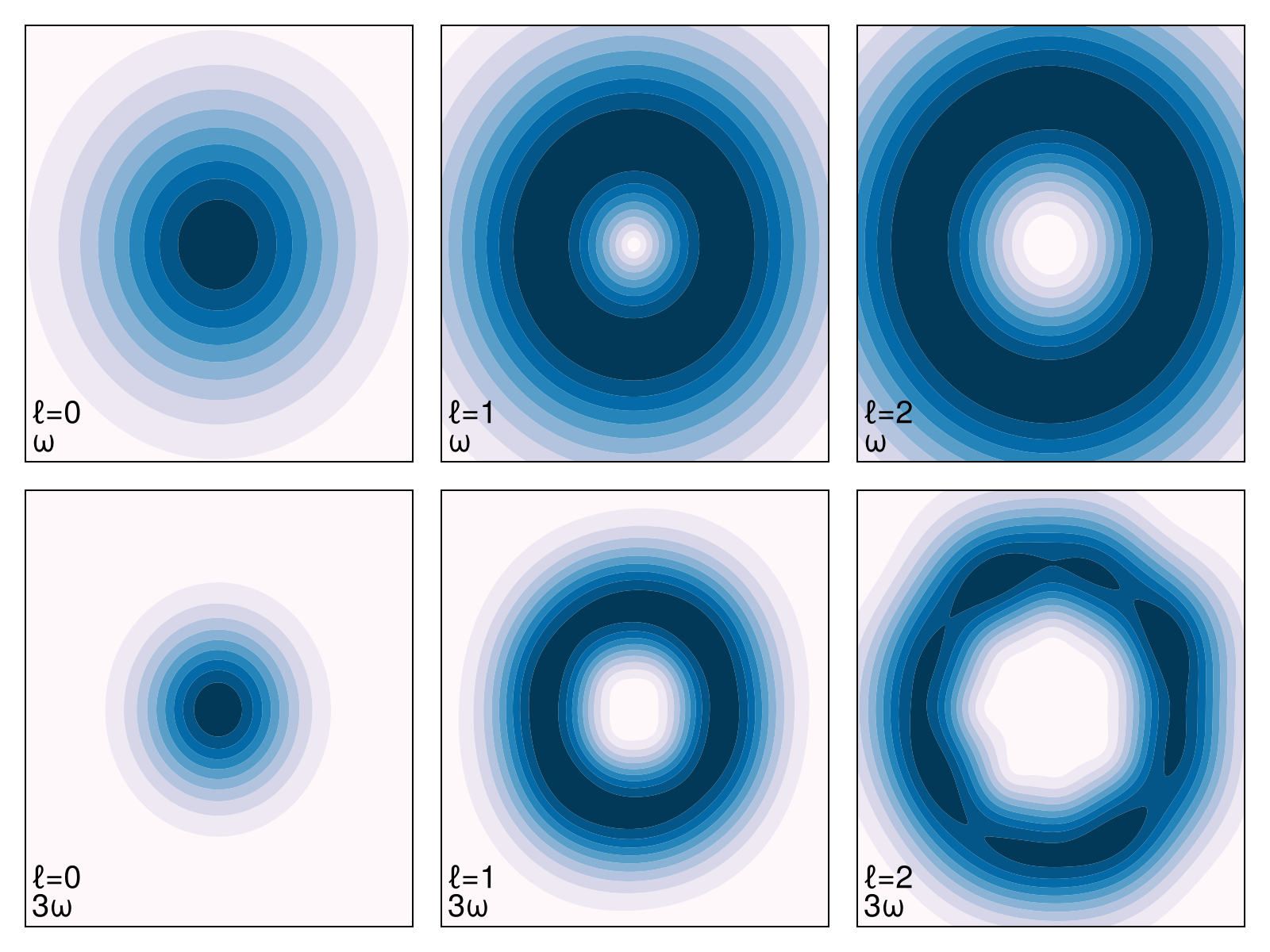}
\caption[width=0.95\textwidth]{Far-field spatial intensity distribution for the fundamental (first row) and third harmonic (second row) components of the nonlinear spectrum $S_{\ell}(\omega)$ generated by the interaction of an OAM pulse with graphene as a function of the OAM charge $\ell$, namely $\ell=0$ (first column), $\ell=1$ (second column), and $\ell=2$ (third column). For $\ell=0$ (first column), we observe the charcacteristic shrinking of the beam waist of the third harmonic beam, compatible with the tripling of the frequency (i.e., wavelength divided by three). For the $\ell\neq 0$ cases, instead, we can see from the far-field intensity distribution that no apparent shrinking of the beam waist occur, but the dimension of the hole in the middle of the vortex beam increases, compatibly with the OAM multiplication law $\ell_n=n\ell_0$, where $\ell_0$ is the OAM carried by the fundamental, impinging beam.}
\label{fig:spatial_distribution}
\end{center}
\end{figure}

\section{Calculations and results}
Without loss of generality, for the rest of the manuscript we assume to consider a linearly polarised electric field, say, along the $x$-direction, and we choose its temporal profile to be Gaussian, such that we can write the vector potential as
\beq\label{eq17}
\vett{A}(\vett{r},t)=A_0e^{-\frac{(t-t_0)^2}{\tau^2}}e^{-i\omega_L t}\,u(\vett{R},z)\uvett{x}+\text{\emph{c.c.}},
\eeq
where $\omega_L$ is the carrier frequency of the pulse, $\tau$ its time duration, $t_0$ is just a time reference, and $u(\vett{R},z)$ is the spatial profile of the pulse, which we assume to be a Laguerre-Gaussian beam, i.e., \cite{OAMbook}
\beq
u(\vett{R},z)=\frac{C_{\ell,p}}{w(z)}\left(\frac{R\sqrt{2}}{w(z)}\right)^{|\ell|}\text{L}_p^{|\ell|}\left(\frac{2R^2}{w^2(z)}\right)e^{i\ell\phi},
\eeq
where $\vett{R}=\{R,\phi\}$, $w(z)=w_0\sqrt{1+(z/z_R)^2}$ is the Gaussian beam waist (with $z_R$ being the Rayleigh range of the LG-beam), $C_{\ell,p}=w_0\sqrt{2p!/[\pi(p+|\ell|)!]}$ is a normalisation factor, and $\text{L}_p^{|\ell|}(x)$ are the associated Laguerre polynomials \cite{nist}, with $\ell\in\mathbb{Z}$ representing the OAM index, and $p\in\mathbb{N}$ representing the radial index of the LG-beam. For the purposes of this work we are only interested in the so-called vortex beams, i.e., LG beams with $p=0$. 

Using the above form for the vector potential, we then solve Eqs. \eqref{dirac_system} numerically using the DifferentialEquations.jl package \cite{rackauckas2017differentialequations} with central difference discretization scheme, and calculate the current according to Eq. \eqref{current}. As parameters for these simulations, for the temporal part we use a central frequency $\omega_L=2\times 10^{15}$ Hz, corresponding to approximately central wavelength of approximately $\lambda_L=1$ $\mu$m, and a time duration for the pulse corresponding to 5 optical cycles, while for the spatial part we set $p=0$ (to consider only vortex modes), and use a beam waist of $w_0=1$ mm at the input plane $z=0$, where the graphene sheet is located. The intensity of the pulse is instead set to $A_0=5\times 10^{-10}$ Wb/m, corresponding to an electric field amplitude of $E_0=10^{5}$ V/m. 

 Once we get the current, we use it as a source term into Maxwell's equations \eqref{eqMaxwell} and use MEEP \cite{oskooi2010meep} to evaluate the nonlinear generated field both in the near- and far-field. This allows us to get complete access to both the spatial and temporal features of the interaction of structured light with graphene.

As an illustrative example, we consider third harmonic generation and look at its features, both in the spatial and frequency domain. First, we look at the spatial intensity distribution in the far field. To do so, we generate the nonlinear fields using Maxwell's equations with the Dirac current as a source, propagate them into the far field using MEEP, and then filter out the third harmonic component to visualise it. The results are presented in Fig. \ref{fig:spatial_distribution}, where, as a function of the OAM charge $\ell$ of the impinging pulse, the fundamental and third harmonic signals are compared. From this figure we can see, that in the case of $\ell=0$, i.e., for a Gaussian beam impinging on graphene, the third harmonic field retains its Gaussian structure, but with a smaller beam waist, in accordance with the fact that the beam waist scales with the harmonic order, i.e., $(w_0)_n=w_0/\sqrt{n}$ in order to keep the Rayleigh range of the beam constant. For $\ell\neq 0$ we do not see this behaviour because the shrinking of the beam waist by a factor $\sqrt{n}$ is compensated by the shifting of the position of the maximum of intensity of the vortex mode, due to the OAM conservation law in harmonic generation, i.e., $\ell_n=n\ell_0$, with $\ell_0$ being the OAM carried by the impinging pulse. To visualise this, let us recall, that the position of the radial maximum of a vortex beam (or, equivalently, the size of the vortex hole) is given by $R=w_0\sqrt{|\ell|/2}$. Therefore, for the third harmonic signal, the position of the radial maximum is found at
\beq
R_3=(w_0)_3\sqrt{|\ell_3|/2}=w_0\sqrt{|\ell/2}, 
\eeq
since the $\sqrt{n}$ shrinking of the beam waist is compensated by the threefold increase of OAM due to third harmonic generation. 

\begin{figure}
\begin{center}
\includegraphics[width=0.45\textwidth]{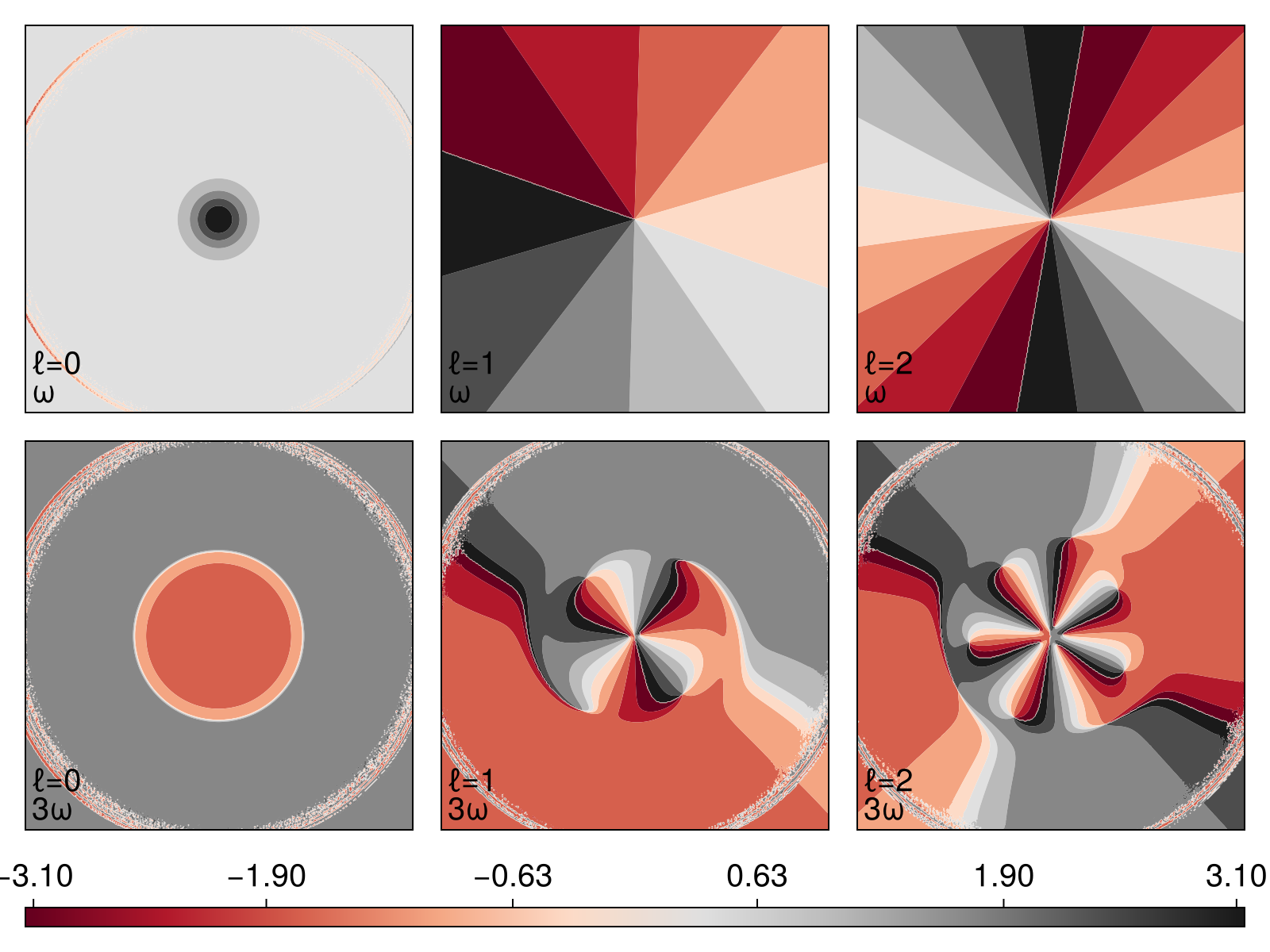}
\caption[width=0.95\textwidth]{Near-field phase distribution for the fundamental (first row) and third harmonic (second row) components of the nonlinear spectrum $S_{\ell}(\omega)$ generated by the interaction of an OAM pulse with graphene, as a function of the OAM charge $\ell$, namely $\ell=0$ (first row, $\ell=1$ (second row), and $\ell=2$ (third row). For $\ell=0$, unsurprisingly, the phase remains flat for both the fundamental and third harmonic signals. For $\ell\neq 0$, in the other hand, the third harmonic signal shows an increase of the OAM charge, according to the OAM conservation law $\ell_3=3\ell_0$, where $\ell_0$ is the OAM charge of the imponging OAM pulse.}
\label{fig:phase_distribution}
\end{center}
\end{figure}

To visualise the OAM conservation law through harmonic generation, it is instructive to look at the near-field phase profile of the fundamental and third harmnonic fields, which are shown in  Fig. \ref{fig:phase_distribution}. Since we are looking at the near-fields, these are essentially equivalent to the phase structure imprinted to 
the Dirac current through the nonlinear interaction of the impinging electromagnetic pulse. As it can be seen from the second and third column the phase profile of the third harmonic indeed contains a topological charge that is three times the one of the impinging beam. This can be proven by counting the black-to-dark-red phase transitions (corresponding to a phase rotation of $2\pi$) at the centre of the phase profile. 

\begin{figure}
\begin{center}
\includegraphics[width=0.45\textwidth]{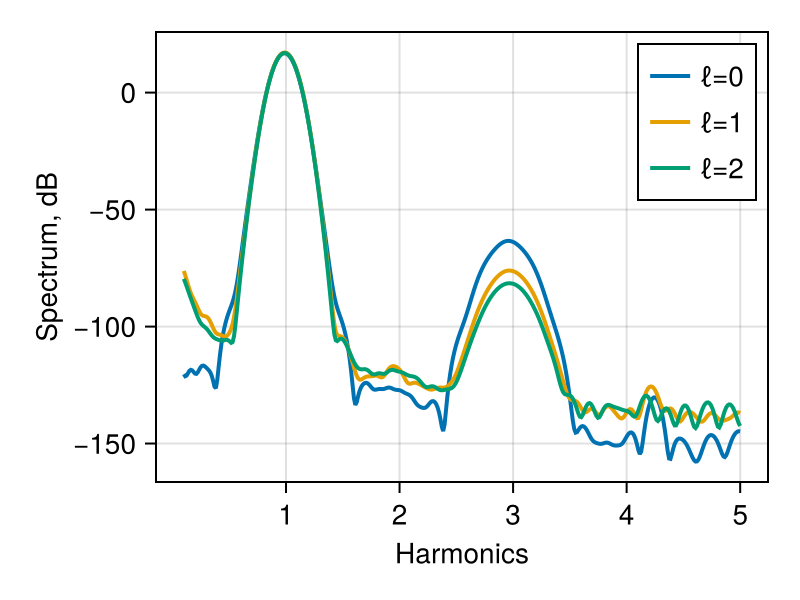}
\caption{OAM-dependent nonlinear Optical Response $S_{\ell}(\omega)$ of graphene, as a function of the different harmonics of the carrier frequency $\omega_L$ of the impinging Laguerre-Gaussian electromagnetic pulse, and for different values of the OAM charge $\ell$, namely $\ell=0$ (blue line), $\ell=1$ (orange line) and $\ell=2$ (green line). As the OAM carried by the impinging pulse increases, the nonlinear response remains structurally the same, but a reduction of the amplitude of the harmonics is observed, due to the peculiar structure of LG beams in $k$-space.}
\label{fig:OAM_spectrum}
\end{center}
\end{figure}

Finally, by integrating with respect to the spatial degrees of freedom, we can calculate the OAM-dependent nonlinear spectrum $S_{\ell}(\omega)$ as 
\beq
S_{\ell}(\omega)=\int\,d^2R\,|\omega\,\tilde{\vett{J}}_{Dirac}(\vett{r},\omega)|^2,
\eeq
whose explicit form as a function of frequency is displayed in Fig. \ref{fig:OAM_spectrum}, limited to the spectral region concerning third harmonic. As it can be seen, the presence of OAM in the impinging beam does not change in a significant manner the structural form of the spectrum, i.e., the second harmonic is still suppressed because the OAM pulse cannot break the inversion symmetry of graphene, given that it itself possesses cylindrical symmetry. However, we observe a reduction of the third harmonic signal with increasing values of the OAM charge $\ell$. This could be due to the fact, that while the angular spectrum of a Gaussian beam ($\ell=0$) is peaked at the position of the Dirac cone and it is therefore capable of generating harmonics from a gapless situation, for OAM modes the maximum of intensity is shifted with respect to the centre of the beam (which is assumed to be at the position of the Dirac cone), and this results in the fact, that the impinging light beam sees, de-facto, a gapped material, since most of the light-matter interaction comes from the region of the Brillouin zone around the maximum of the LG mode. This, ultimately, reduces the efficiency of third harmonic generation, since for higher OAM values, electrons near a Dirac point behave like a usual gapped material, rather than seeing a nearly zero gap, as in the case of a Gaussian beam. 

\section{Conclusion}\label{sec:conc}
In this work, we introduced a framework to study the light-matter interaction of graphene with structured light and analyzed the optical response in the nonlinear regime. Our model is based on the tight-binding, low-energy approximation of the graphene Hamiltonian, which can be represented as a massless Dirac equation in the vicinity of the Dirac cones. We have coupled this with a spatially-varying electromagnetic pulse and analysed the generation of harmonics as a function of of the OAM parameter $\ell$ of the impinging pulse. By coupling back the Dirac equation with Maxwell's equations, through the generated nonlinear current, we were able to get information on both the frequency-resolved intensity and phase distribution of the emitted nonlinear signal. As an example of application of our framework, we have considered the explicit example of third harmonic generation and studied both the intensity and phase profile of the generated third harmonic field, as well as the nonlinear spectrum of harmonics as a function of the OAM charge $\ell$.

One clear advantage of our model is the possibilty to avoid describing structured light-matter interaction in $k-$space (a task that proves to be quite challenging, both from an analytical and computational perspective \cite{tamashevich2022inhomogeneous}, and focus only on direct space. By considering the full coupling between graphene and light, i.e., including light-matter interaction in the low energy Hamiltonian of graphene, and the generated current as source of Maxwell's equation, our approach is capable of providing versatile access to different observables and properties of the field, allowing access to both the near- and far-field, for frequency generated by the interaction. 

Finally, although in this work we have focused our attention to graphene, our results can be easily adapted to other 2D material, and extended to include different situations, like external magnetic fields, or specific geometries. This, ultimately, has the potential to  provide a general framework to include the effects of structured light in both the linear and nonlinear optical response of such materials.

\section*{Acknowledgements}
We acknowledge the financial support of the Academy of Finland Flagship Programme, Photonics research and innovation (PREIN), decision 346511. Y.T. also acknowledges support from the Finnish Cultural Foundation, decision 00231133.

\bibliography{Bibliography}.

\end{document}